# An Abstract Method Linearization for Detecting Source Code Plagiarism in Object-Oriented Environment

Oscar Karnalim
Faculty of Information Technology
Maranatha Christian University
Bandung, Indonesia
oscar.karnalim@it.maranatha.edu

*Abstract*—**Despite the fact that plagiarizing source code is a trivial task for most CS students, detecting such unethical behavior requires a considerable amount of effort. Thus, several plagiarism detection systems were developed to handle such issue. This paper extends Karnalim's work, a low-level approach for detecting Java source code plagiarism, by incorporating abstract method linearization. Such extension is incorporated to enhance the accuracy of low-level approach in term of detecting plagiarism in object-oriented environment. According to our evaluation, which was conducted based on 23 design-pattern source code pairs, our extended low-level approach is more effective than state-of-the-art and Karnalim's approach. On the one hand, when compared to state-of-the-art approach, our approach can generate less coincidental similarities and provide more accurate result. On the other hand, when compared to Karnalim's approach, our approach, at some extent, can generate higher similarity when simple abstract method invocation is incorporated.**

## I. INTRODUCTION

In Computer Science (CS) major, conducting plagiarism is a trivial task since most students are familiar with copy-and-paste technique. They could copy and plagiarize their colleague's work in no time. In addition, since most assignments are written electronically in such major, most students could disguise their unethical behavior without leaving a particular trace. As we know, the modification of most electronic files would replace the original one in exact position with no trace. In order to handle such issue, numerous plagiarism detection systems are developed where most of them are focused on programming assignment plagiarism.

In term of detecting plagiarism in programming assignment, most works are classified into two categories which are attribute- and structure-based approach [1]. Attribute-based approach detects plagiarism based on key properties from given source codes whereas structure-based approach detects plagiarism based on source code ordinal structure. According to several works [2, 3], the latter one frequently outperforms the first approach in term of its effectiveness. Such approach is more resistible to handle high-level plagiarism attack since it considers the structural order of compared source code.

This paper extends Karnalim's work [4], a structure-based approach for detecting source code plagiarism, with an additional mechanism, namely abstract method linearization. Such mechanism is intended to naively linearize the content of abstract methods so that low-level approach could become more accurate for detecting plagiarism in object-oriented environment. Our proposed approach is then evaluated based on 23 design-pattern source codes with its respective anti-pattern as its plagiarized code. Such dataset is assumed to represent source codes that are implemented in complex object-oriented environment.

## II. RELATED WORKS

When detecting source code plagiarism, structure-based approach uses ordinal structure from source code to determine similarity value. Such approach usually works by translating source codes into their respective intermediate form and compare such forms in pairwise manner through similarity algorithm [1]. In general, there are three intermediate representations which are frequently used in structure-based approach. These representations are lexical token sequence [3], compiler-based representation [5], or low-level codes [6, 7, 8, 4, 9]. Among these forms, low-level code, at some extent, is claimed to be quite effective and efficient due to its token compactness [9, 4]. It only contains semantic-preserving instructions and most syntactic sugars are automatically translated into its original form.

To the best of our knowledge, there are five works which incorporated low-level codes for detecting source code plagiarism. They are classified into two categories based on their target programming languages. On the one hand, three of them are focused on .NET programming language. It was initiated by Juričić's work [6] which detect plagiarism based on Levenstein distance of Common Intermediate Language instructions. That work was then followed by Juričić et al [7] and Rabbani & Karnalim [9] by replacing Levenstein distance with more-sophisticated similarity measurement. On the other hand, two of them are focused on Java programming language. It was initiated by Ji et al's work [8] which converts Java source code into Bytecode and compare them through Adaptive Local Alignment algorithm. Such work was then expanded by Karnalim [4] by replacing similarity algorithm with RKGST and incorporating supplementary mechanisms, such as recursion-handling method linearization, instruction generalization, and instruction interpretation.

In this paper, an expanded version of Karnalim's work [4] is presented for handling source code plagiarism in object-oriented environment. Our work incorporates abstract method linearization, a mechanism to naively detect the content of abstract method without executing given code directly. Such mechanism is expected to linearize abstract method based on its implementer contents instead of replacing it with an empty string during method linearization.

### III. OVERVIEW OF LOW-LEVEL APPROACH

In general, our low-level approach, which is extended from Karnalim's work [4], works in threefold. Firstly, all target source codes will be compiled into their respective low-level form. In our case, all Java source code will be compiled into their respective class files which contain bytecode instructions. Secondly, generated class files are then extracted to readable low-level token sequences. It is important to note that extraction phase conducted in this work is expanded with abstract method linearization. Finally, all generated low-level token sequences will be compared to each other using local RKGST minimum matching similarity with 2 as RKGST minimum matching threshold. Such mechanism deducts similarity based on the collective matched tokens from all method pairs, where the pairing mechanism is based on method signature similarity.

Abstract method linearization intends to approximately linearize abstract method content based on their respective implementer methods. All tokens from implementer methods will be concatenated and reconsidered as the content of given abstract method. In such manner, the content of abstract method is still able to be approximately predicted even though its method is invoked through generalized-type variable. In term of its implementation, abstract method linearization will be conducted right before standard method linearization from extraction phase in Karnalim's work [4]. It is conducted into three steps. Firstly, all class and interface relations would be remodeled as a graph, where A → B represents that class/interface A implements or inherits B. Secondly, all nodes in generated graph are then sorted based on incoming edges in ascending order through topological sort [10]. Finally, all abstract methods for each node would be linearized sequentially, started from node with the lowest number of incoming edges.

For more details about how our abstract method linearization works, let see a case study with sample class structure given on Figure 1. Initially, all class and interface relations are converted into a graph and sorted in ascending order using topological sort. This phase sorts classes and interfaces into *ConcreteClass1*, *ConcreteClass2*, *ConcreteClass3*, *ConcreteClass4*, *ConcreteClass5*, *Interface2*, *AbstractClass2*, *Interface1*, and *AbstractClass1*. Since the first five classes have no abstract methods, we will start to discuss linearization process from the 6th to the 9th element. Firstly, all methods from *Interface2* are linearized based on *ConcreteClass5* methods which have similar method signature since *ConcreteClass5* is the only implementer of that interface. Both methods will yield exactly similar content with its implementer method since such interface is only implemented once. Secondly, *AbstractClass2* has 2 abstract methods which are *foo1* and *foo2*. Each method is then linearized by concatenating the content of similar method from *ConcreteClass2* and *ConcreteClass3*. In other words, each abstract method from *AbstractClass2* will have more tokens since they combine the content of two implementer methods. Thirdly, *Interface1*'s *foo3* is linearized based on the concatenation of methods which have similar signature from *ConcreteClass4* and *Interface2*. It is important to note that *foo3* on *Interface2*, which is required to linearize *Interface1*'s *foo3*, had been linearized in previous iteration. By using topological sort, our approach guarantees that all required abstract methods have been linearized before used. Finally, *foo1* in *AbstractClass1* is linearized based on similar-signatured method from *ConcreteClass1* and *AbstractClass2*. Even though *AbstractClass2*'s *foo1* is abstract, its content has been linearized before on previous iteration. Therefore, *foo1* on *AbstractClass1* could be linearized by simply concatenating the content of both methods.

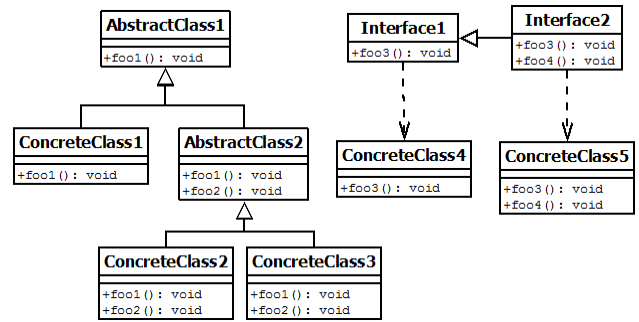

Figure 1. Sample Class Structure

### IV. EVALUATION

This evaluation is conducted to measure how far object-oriented environment affects resulted similarity of low-level approach. In order to do that, 23 design patterns, which were defined by Gang of Four [11], are selected as our case study. For each design pattern, original source code would be taken from [12] whereas its plagiarized code would be generated by removing such pattern on given code. In such manner, the original and plagiarized code are only different in term of incorporated design pattern. We do believe that the existence of such pattern represents object-oriented environment since most design patterns are formed by incorporating complex object-oriented techniques. The detail of involved design patterns with its unique ID, which will be used for convenient reference at the rest of this paper, can be seen on Figure 2. These cases are sorted per pattern category where case 1-5 refer to creational patterns; case 6-16 refer to behavioral patterns; and case 17-23 refer to structural patterns.

This evaluation incorporates 3 scenarios which are low-level approach (LA), low-level approach without abstract method linearization (LA-M), and standard lexical token approach (SLT). Firstly, LA refers to our proposed mechanism, that is featured with abstract method linearization. Secondly, LA-M refers to our proposed mechanism without the existence of abstract method linearization. This scenario will be used to measure the impact of abstract method linearization by comparing its result with LA. Finally, SLT refers to state-of-the-art approach that converts source codes into lexical token streams and compares them using RKGST algorithm with 2 as its minimum matching threshold. This scenario will be used as a baseline to measure the effectiveness improvement provided by low-level approach.

| ID | Design Pattern | ID | Design Pattern | ID | Design Pattern |
|---|---|---|---|---|---|
| C01 | Abstract Factory | C09 | Iterator | C17 | Adapter |
| C02 | Builder | C10 | Mediator | C18 | Bridge |
| C03 | Factory | C11 | Memento | C19 | Composite |
| C04 | Prototype | C12 | Observer | C20 | Decorator |
| C05 | Singleton | C13 | State | C21 | Facade |
| C06 | Chain of Responsibility | C14 | Strategy | C22 | Flyweight |
| C07 | Command | C15 | Template | C23 | Proxy |
| C08 | Interpreter | C16 | Visitor | | |

Figure 2. Evaluation Dataset with its Respective ID

Since low-level form is more compact than the source code itself [4, 9], comparing the result of these scenarios based on normalized similarity might be unfair. One mismatch token on low-level approach may significantly lower its normalized similarity due to its small token size. Thus, a measurement without normalization, namely Inverse number of Mismatched Token (IMT), is proposed. The detail of such measurement can be seen in (1). As its name states, it is inversely proportional to the number of Mismatched Token (MT). MT, which is adapted from Karnalim's evaluation scenario [4], is generated by subtracting the number of involved tokens with the number of matched tokens.

$$IMT(A) = -1 * MT \qquad (1)$$

The detail of our evaluation result can be seen on Figure 3 where vertical axis represents IMT value for each case and horizontal axis represents plagiarism cases in our evaluation dataset. In general, SLT yielded the most fluctuated result when compared to other approaches. It generates the highest IMT on several cases while resulting the lowest one on the others. On the other hand, the result of LA and LA-M are less fluctuated. They generate narrower result range when compared to SLT.

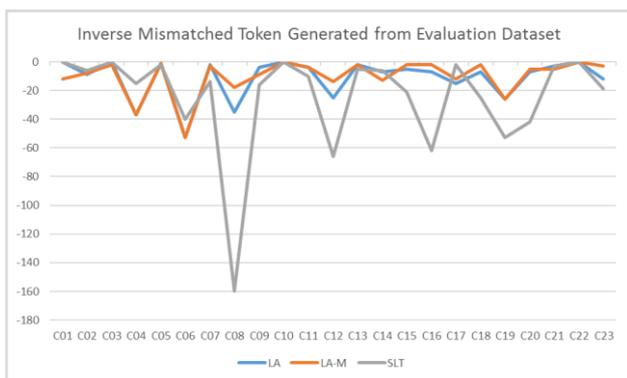

Figure 3. Evaluation Result

According to Figure 3, the highest IMT, which is 0, is occurred ten times where four of them were generated as LA results; two of them were generated as LA-M results; and the others were generated as SLT results. These cases yield high IMT since incorporated plagiarism attacks on such cases are perfectly matched with scenario characteristics. On the contrary, the lowest IMT is generated by SLT when detecting similarity on C08 case. Such finding is natural since removing interpreter pattern in C08 generates a significant modification in source code level and all tokens involved on such modification were considered as mismatched tokens by SLT.

LA has several major benefits when compared to SLT in term of effectiveness. Firstly, LA generates less coincidental similarities since such approach considers program context when determining similarity. It only compares subsequences from methods which share considerably similar signature and excludes delimiter tokens from its comparison due to the nature of low-level form characteristics. It is quite different with SLT which treats the whole source code as a long sequence and includes all delimiter tokens on its comparison. Similar subsequences could be found anywhere without considering given context and similarity result might be biased due to the inclusion of delimiter tokens. Such benefit is deducted from C02, C04, C06, C14, and C17 where SLT seems to outperform LA. Secondly, LA result would be more semantically-related since such approach only compares semantic-preserving tokens. No delimiter tokens would be involved in low-level form since they have been removed at compilation phase. Such benefit is deducted from C09, C12, and C19 where LA outperforms SLT due to less mismatched tokens. Thirdly, LA would be more compatible for detecting student source code plagiarism since it considers standard programming mechanisms such as method invocation and abstract method. These mechanisms are frequently used by students on programming task, especially in object-oriented environment. Such benefit is deducted from C15 and C23 where SLT yields lower IMT when compared to LA, which, at some extent, considers such programming mechanism. Finally, LA enables plagiarism detection to be measured dynamically per method thanks to method pair mechanism. Such phenomenon is beneficial in term of excluding more mismatched tokens since minimum matching similarity could dynamically replace tokens that act as a baseline for determining similarity. In addition, since method pair mechanism is only focused on method content similarity, class and attribute modification are unavailing on such approaches. Such benefit is deducted from C07, C08, C12, C13, C16, C18, C20, and C23 where SLT yields low IMT on these cases due to static minimum mechanism.

Despite their benefits, LA has an additional drawback, which is mainly caused by the compactness of low-level form. At some points, it cannot detect similar short subsequence which can be detected through SLT. This finding is natural since source code tokens are less concise than low-level tokens. In low-level form, several source code tokens are replaced with less tokens that refer to the same semantic. Thus, when both SLT and LA incorporate similar minimum matching threshold, SLT could detect more matched tokens due to the longer representation of source code tokens. Such drawback is deducted from C02 where LA cannot generate higher IMT than SLT since the length of several matched subsequences is lower than incorporated RKGST minimum matching threshold, which is 2. At that case, it is important to note that reducing such threshold to 1 is not a solution since the number of false-positive tokens would be increased.

Abstract method linearization yields adequate impact since it enhances IMT on several cases while reducing IMT on other cases. Such linearization is effective for handling abstract method that only has one implementer. This benefit can be explicitly seen on C01, C03, and C09 where LA yields higher IMT than LA-M. However, such linearization is less effective when handling abstract method with numerous implementers. It typically generates more mismatched tokens than the matched ones since it concatenates all possible linearization as method content and only a small portion of them which are considered as actual matches. This drawback can be explicitly seen on C08, C15, C16, C18, and C23 where LA yields lower IMT than LA-M.

Based on our manual observation toward evaluation dataset, standard method linearization, which has been initially developed by Karnalim [4], takes a significant role for determining similarity in object-oriented environment. Most cases incorporated method inlining and outlining when generating the plagiarized code. Thus, without method linearization, such cases would not be detected correctly since numerous method content will be ignored during comparison phase.

## V. Conclusion

In this paper, a low-level approach, which is extended from Karnalim's work [4], is proposed. It incorporates abstract method linearization to predict the content of abstract method by concatenating all method contents from its respective implementer(s). Based on our dataset, our low-level approach outperforms state-of-the-art approach in term of its sensitivity. It could generate less coincidental similarities and provide more accurate result since it only considers semantic-preserving tokens. Furthermore, such approach is more suitable to handle student source code plagiarism since it considers several programming concepts and detects similarity based on method content instead of the whole tokens. Despite the benefits, our approach suffers a drawback. It cannot detect several short similar pairs which can be detected through standard lexical token approach. Such issue is caused by the compactness of low-level tokens where several source code tokens are compressed into less tokens in low-level form. However, since the benefits outweigh the drawback, we would argue that our approach is quite promising to detect source code plagiarism in object-oriented environment.

Abstract method linearization is proved to be considerably effective for detecting plagiarism in object-oriented environment, especially when the number of its implementer is considerably small. It can generate method content of invoked abstract method without directly executing the codes. Yet, it would generate numerous mismatched tokens when given abstract methods have been reimplemented many times.

For future work, our proposed method will be combined with attribute-based approach so that it could detect source code plagiarism in more-effective manner. In addition, we also intend to evaluate our approach on real plagiarism cases that incorporate object-oriented paradigm. We want to revalidate whether our hypothesis about its effectiveness is also applied to real plagiarism cases or not.

## VI. Threats to Validity

It is important to note that evaluation dataset incorporated in our work does not represent all possible implementations of 23 design patterns mentioned by Gamma et al [11]. Such evaluation dataset only represents a subset of them. Thus, the results provided in this work is not guaranteed to represent the impact of our approach for given design patterns in general. We only used such dataset as a representation of source code in complex object-oriented environment. In addition, due to various object-oriented techniques, our findings cannot be generalized to all possible combinations of object-oriented techniques. We tried to mitigate this issue by incorporating various design patterns as plagiarism attacks. Yet, it might only represent a small subset of possible combined techniques in object-oriented environment.

## References

[1] Z. A. Al-Khanjari, J. A. Fiadhi, R. A. Al-Hinai and N. S. Kutti, "PlagDetect: a Java programming plagiarism detection tool," in *ACM Inroads*, New York, ACM, 2010, pp. 66-71.

[2] L. Prechelt, G. Malpohl and M. Philippsen, "Finding plagiarisms among a set of programs with JPlag," *Journal of Universal Computer Science,* vol. 8, no. 11, 2002.

[3] Z. Djuric and D. Gasevic, "A Source Code Similarity System for Plagiarism Detection," *The Computer Journal,* vol. 55, 2012.

[4] O. Karnalim, "Detecting Source Code Plagiarism on Introductory Programming Course Assignments Using a Bytecode Approach," in *The 10th International Conference on Information & Communication Technology and Systems (ICTS)*, Surabaya, 2016.

[5] M. Chilowicz, E. Duris and G. Roussel, "Syntax tree fingerprinting for source code similarity detection," in *IEEE 17th International Conference on Program Comprehension*, Vancouver, 2009.

[6] V. Juričić, "Detecting source code similarity using low-level languages," in *33rd International Conference on Information Technology Interfaces*, Dubrovnik, 2011.

[7] V. Juricic, T. Juric and M. Tkalec, "Performance evaluation of plagiarism detection method based on the intermediate language," *INFuture2011: "Information Sciences and e-Society",* 2011.

[8] J.-H. Ji, G. Woo and H.-G. Cho, "A Plagiarism Detection Technique for Java Program Using Bytecode Analysis," in *ICCIT '08. Third International Conference on Convergence and Hybrid Information Technology*, Busan, 2008.

[9] F. S. Rabbani and O. Karnalim, "Detecting Source Code Plagiarism on .NET Programming Languages using Low-level Representation and Adaptive Local Alignment," *Journal of Information and Organizational Sciences,* vol. 41, no. 1, 2017.

[10] R. Sedgewick and K. Wayne, Algorithms, 4th Edition, Addison-Wesley, 2011.

[11] E. Gamma, R. Helm, R. Johnson and J. Vlissides, Design Patterns: Elements of Reusable Object-Oriented Software, USA: Addison-Wesley, 1994.

[12] "Design Patterns in Java Tutorial," [Online]. Available: https://www.tutorialspoint.com/design_pattern/. [Accessed 23 2 2017].